\documentclass{jpp}
\usepackage{graphicx}
\usepackage{epstopdf, epsfig}

\newcommand{\be}{\begin{equation}}
\newcommand{\ee}{\end{equation}}
\newcommand{\bea}{\begin{eqnarray}}
\newcommand{\eea}{\end{eqnarray}}
\newcommand{\lang}{\left\langle}
\newcommand{\rang}{\right\rangle}
\newcommand{\bs}{\boldsymbol}

\shorttitle{Impurity transport in a mixed collisionality stellarator plasma}
\shortauthor{S. L. Newton, P. Helander, A. Moll\'{e}n and H. M. Smith}

\title{Impurity transport and bulk ion flow in a mixed collisionality stellarator plasma}

\author{S. L. Newton\aff{1},\aff{2}
  \corresp{\email{sarahn@chalmers.se}},
	P. Helander\aff{3}
  A. Moll\'{e}n\aff{3}
 \and H. M. Smith\aff{3}}

\affiliation{
\aff{1}Department of Physics, Chalmers University of Technology, G\"{o}teborg SE-412 96, Sweden
\aff{2}CCFE, Culham Science Centre, Abingdon, Oxon OX14 3DB, UK
\aff{3}Max-Planck-Institut f\"{u}r Plasmaphysik, 17491 Greifswald, Germany}

\begin{document}

\maketitle

\begin{abstract}
The accumulation of impurities in the core of magnetically confined plasmas, resulting from standard collisional transport mechanisms, is a known threat to their performance as fusion energy sources. 
Whilst the axisymmetric tokamak systems have been shown to benefit from the effect of temperature screening, that is an outward flux of impurities driven by the temperature gradient, impurity accumulation in stellarators was thought to be inevitable, driven robustly by the inward pointing electric field characteristic of hot fusion plasmas.
We have shown in~\citet{helanderetal2017b} that such screening can in principle also appear in stellarators, in the experimentally relevant mixed collisionality regime, where a highly collisional impurity species is present in a low collisionality bulk plasma. Details of the analytic calculation are presented here, along with the effect of the impurity on the bulk ion flow, which will ultimately affect the bulk contribution to the bootstrap current.
\end{abstract}


\section{Introduction}
\label{secintro}

Magnetic confinement fusion requires a plasma to be maintained at multi-keV temperatures in near steady state conditions. 
The two leading types of device used to achieve this are the axisymmetric tokamak and non-axisymmetric stellarator.
Whilst they have a number of competing advantages and disadvantages~\citep{helanderetal2012}, which are still being studied and mitigation techniques developed, both suffer from the potential threat of accumulation of impurities in the hot core plasma~\citep{connor1973,hirschetal2008}.
Released during plasma-wall interactions, impurities can make their way into the confined bulk plasma.
Precautions are taken to minimise dilution of the plasma (which would reduce the fusion reactivity) by the choice of low atomic number materials for the walls of the device, but typically heavy materials must be chosen for the plasma exhaust region~\citep{joffrinetal2014}. Heavy impurities are not fully ionised at typical operating temperatures, and power balance cannot be maintained in the presence of the radiation emitted by a significant accumulation, so the plasma would quench.
Therefore, the behaviour of impurity ions in hydrogen-isotope plasmas must be understood, to ensure that it can be controlled.

Particle transport in magnetically confined plasmas results from both turbulent and neoclassical processes. The latter is essentially a random walk due to collisions between particles as they move along the variety of trajectories set by the magnetic field structure.
Turbulent transport dominates many aspects of confined plasma behaviour, but for heavy impurity ions the neoclassical transport is known to be significant, in both tokamaks and stellarators, with the bulk ion density gradient producing a strong inward flux, and so, impurity accumulation~\citep{angionihelander2014,hirschetal2008}. 
However, in tokamaks, the velocity dependence of the inter-species collision frequency is known to lead to an impurity flux driven by the bulk ion temperature gradient, whose sign depends on the collisionality regime of the bulk ions~\citep{connor1973,hirshman1977}. Denoted by $\nu_{*ab}$ for collisions between species $a$ and $b$ (and defined in detail in section~\ref{secfluxfrictiondef}), the collisionality represents the ratio of the typical size of the device to the particle mean free path. When the bulk ions (denoted throughout by $i$) are in the low collisionality regime, $\nu_{*ib} < 1$, an outward impurity flux is driven by the temperature gradient. This ``temperature screening'' was identified experimentally in~\citet{wadeetal2000}. Whilst the temperature gradient typically drives an inward flux when the bulk ions are in the high collisionality regime, it was noted that an outward flux could still be driven in rather clean plasmas~\citep{rutherford1974}.

Importantly, in tokamaks the net transport driven by the radial electric field vanishes. This is not the case in a stellarator, where not only does a net particle flux result from the radial electric field, it is proportional to the particle charge -- and therefore this contribution is usually expected to dominate the transport of heavy impurities. With the radial electric field in hot stellarator plasmas typically pointing inward~\citep{hirschetal2008,klingeretal2017}, a large inward flux arises and the picture of impurity transport in stellarators appears bleak~\citep{hirschetal2008,velascoetal2017}. Accumulation is indeed often seen experimentally~\citep{w7a1985,igitkhanovetal2006,hirschetal2008}, although the exceptional behaviour of low-density ``impurity-hole'' plasmas in LHD is still to be understood~\citep{idaetal2009}.
Yet, such conclusions were primarily based on calculations in which the collision operator describing inter-species collisions was approximated by a scattering operator, accounting for the deflection of the particle pitch-angle with respect to the magnetic field line, sometimes including an additional term to ensure momentum conservation. Numerical codes retaining only scattering interactions between species have been routinely used to calculate stellarator neoclassical transport~\citep{beidleretal2011}. However it is known that such operators cannot correctly treat high collisionality species, and hence the experimentally relevant mixed collisionality regime, where a heavy, highly charged and thus collisional impurity species (denoted here by a subscript $z$, with charge $Ze$) is present in a low collisionality hydrogenic bulk plasma.

We have therefore calculated the impurity flux across nested magnetic flux surfaces in such a mixed collisionality plasma analytically. A summary of the results appeared in~\citet{helanderetal2017b}, along with an initial successful comparison to the numerical results from the drift-kinetic equation solver SFINCS~\citep{landremanetal2014}, which retains the full linearised Landau collision operator and can treat multiple species. Here we provide full details of the analytic calculation, whilst a more extensive numerical comparison will appear separately. 
The complicated stellarator field structure means that the bulk ions can exist in a series of low collisionality regimes, unlike a tokamak plasma. We have treated both the moderate $1/\nu$ collisionality regime, where the radial drift of particles trapped in localised magnetic wells is interrupted sufficiently frequently by collisions to prevent direct loss of particles from the plasma, and the lower collisionality $\sqrt{\nu}$ regime, where magnetic field optimisation, or the averaging effect of the drift within the flux surfaces produced by a sufficiently strong radial electric field, is required to ensure good confinement. The transport of impurities in a highly collisional stellarator, applicable to the cooler edge plasma, was studied analytically by~\citet{braunhelander2010}, and we adopt the same flux-friction formalism.
We also present a short extension, giving the cross-field flux of the heaviest impurity when two collisional impurity species, of disparate mass, are present in the low collisionality bulk. This may be of particular relevance experimentally, where heavy impurities from exhaust components, such as Fe or potentially W in future devices, are often present in small quantities in a main H bulk plasma, with another dominant, but lighter impurity, released from the main plasma facing components.

Finally, note that the confining magnetic field in a stellarator is primarily produced by external coils~\citep{landreman2017}, and in the design of a stellarator, a numerical optimisation process of coil positioning and current values is typically undertaken.
Beside cross-field transport, another important neoclassical effect in an inhomogeneous plasma is the self-generated bootstrap current. In a tokamak this helps to maintain the current needed to confine the plasma, but in a stellarator it can distort the confining field and may have to be minimised~\citep{geigeretal2015}.
The bulk ion flow and bootstrap current were recently determined analytically for a pure plasma, in which the bulk ions were taken to be in the $1/\nu$ or $\sqrt{\nu}$ collisionality regimes~\citep{helanderetal2017a}.
As the plasma flow naturally follows from the flux-friction formalism we also determine the effect of the impurity on the bulk ion flow here, which will affect the final bootstrap current. 

The paper is organised as follows.
In section~\ref{secformalism}, we outline the flux-friction formulation for the impurity flux, and present the solution for the species' distribution functions in the different collisionality regimes, using model collision operators at low collisionality.
The radial impurity flux is then evaluated in section~\ref{secflux} and expressed in terms of transport coefficients, which give the response of the flux to the various driving gradients. The impurity content appears only as a prefactor in the impurity flux.
When the bulk ions are in the $1/\nu$ regime, the structure of the impurity flux is similar to the high collisionality case, with the transport produced by the impurity and bulk ion density gradients simply related by the impurity charge.
With the bulk ions in the $\sqrt{\nu}$ regime, additional geometry factors appear in the coefficients relating the impurity flux to the bulk ion gradients.
We find that temperature screening is possible in both of the low collisionality regimes. As mentioned above this is contrary to the usual expectation. Furthermore, we see that the drive from the radial electric field vanishes when the bulk is in the $1/\nu$ regime, and can remain weak into the $\sqrt{\nu}$ regime, under certain conditions.
In section~\ref{secflow} we determine the bulk ion flow in the presence of impurities, again expressing this in terms of transport coefficients, which are sensitive to the impurity content.
We conclude with a discussion in section~\ref{secdiscussion}.


\section{Formulation}
\label{secformalism}

The neoclassical impurity flux can be conveniently expressed in the following form~\citep{igitkhanovetal2006,helanderetal2012}
\be
\Gamma_z = \lang \int f_z ({\bs v}_{dz} \cdot \nabla r) d^3v \rang
= n_z \left( D_{11}^{zi} A_{1i} + D_{11}^{zz} A_{1z} + D_{12}^{z} A_{2i} \right),
\label{eqdeffluxcoeffs}
\ee
where the set of transport coefficients $D$ relate the flux across the magnetic surfaces to the various driving ``thermodynamic forces'', with a prime denoting the derivative with respect to the argument,
$$
A_{1a} = \frac{d \ln p_a}{d r} + \frac{e_a \Phi'(r)}{T_a}, \qquad A_{2a} = \frac{d \ln T_a}{d r}.
$$
Here $p_a = n_a T_a$ is the pressure of species $a$, with charge $e_a$ and $\Phi(r)$ is the electrostatic potential. Stellarator geometry precludes rapid toroidal rotation in general, and along with the density $n_a$ and temperature $T_a$, the potential is approximately constant on magnetic surfaces, which is discussed further below. (The electric field parallel to the magnetic field is here taken to be negligibly weak, but this is not always a good approximation in a stellarator, as shown in~\citet{garciareganaetal2017}.) We also assume that the plasma is sufficiently well confined (see section~\ref{secbulkdistrib}) that the temperatures of the ion species have equalised $T_z = T_i = T$, and so $A_{2z} = A_{2i}$. 

The nested magnetic flux surfaces are labelled by $r$, which acts as an arbitrary radial coordinate, and the angular brackets indicate the average over a flux surface.
Finally we note that the drift velocity of a species, ${\bs v}_{da}$, can usefully be written in conservative form
\be
{\bs v}_{da} = \frac{v_\parallel}{\Omega_a} \nabla \times \left(v_\parallel {\bs b}\right),
\label{eqconsvd}
\ee
where ${\bs b} = {\bs B} / B$, ${\bs v}$ is the particle velocity, $\Omega_a = e_a B /m_a$ is the gyrofrequency for a species with mass $m_a$, parallel and perpendicular are taken throughout with respect to the magnetic field ${\bs B}$, and the curl is taken at constant particle energy $\epsilon_a = m_a v^2 / 2 + e_a\Phi$ and magnetic moment $\mu_a = m_a v_\perp^2 / 2B$.
In the following subsections we describe the formalism used to calculate the radial impurity flux, and hence the transport coefficients, which we present in section~\ref{secflux}.


\subsection{Flux-friction relation}
\label{secfluxfrictiondef}

The formulation of the radial impurity flux in a stellarator in terms of a flux-friction relation was detailed in~\citet{braunhelander2010,sugamanishimura2002}.
The flux is decomposed into a sum of contributions, the first due to friction against the background bulk ions and the second the result of the impurity pressure anisotropy,
\be
\Gamma_z = \lang \int f_z ({\bs v}_{dz} \cdot \nabla r) d^3v \rang = \frac{1}{Z e}\lang uBR_{z\parallel} + \left(p_{z\parallel} - p_{z\perp}\right)\frac{\nabla_\parallel(uB^2)}{2B}\rang.
\label{eqgenradialflux}
\ee
The effect of friction against electrons is small in the electron-ion mass ratio, so it is neglected throughout.
The equilibrium function $u$ satisfies ${\bs b} \cdot \nabla u = - {\bs b} \times \nabla r \cdot \nabla (B^{-2})$.
With the linearised, gyroaveraged, collision operator for species $a$ denoted by $C_a = \sum_{b} C_{ab}$, where the sum is over the ion species present, we can compare the magnitude of the flux driven by the parallel friction $R_{z\parallel} = m_z \int v_\parallel C_z(f_{z}) d^3v$ to that expected due to the species' pressure anisotropy $p_{z\parallel} - p_{z\perp}$, by considering the first terms in an expansion of the drift kinetic equation governing the impurity behaviour. 

The expansion is taken as usual with respect to the magnetisation parameter $\rho_{*z} = \rho_z/L$~\citep{helander2014}, where $\rho_a$ is the gyroradius of species $a$, and $L$ is a characteristic length scale perpendicular to the background magnetic field. We assume $Z \gg 1$, but not so large to require that $\rho_{*z}$ is higher order with respect to $\rho_{*i}$.
Taking a characteristic parallel length scale $L_\parallel$, which will satisfy $L_\parallel > L$, the ratio of the contributions to the flux in eq.~(\ref{eqgenradialflux}) is approximately $\left(p_{z\parallel} - p_{z\perp}\right)/R_{z\parallel}L_\parallel$.
Due to the high collisionality the leading order piece of the expanded distribution function $f_{z} = f_{z0} + f_{z1} + \ldots$ will be a Maxwellian, $f_{Mz} = \left(n_z/\pi^{3/2}v_{Tz}^3\right)  \exp \left( - v^2/v_{Tz}^2\right)$, where the thermal velocity of a species is $v_{Ta} = \sqrt{2T_{a}/m_a}$.
The first order drift kinetic equation for the distribution function $f_{z1}$ then takes the form
\be
C_z \left(f_{z1}\right) = v_\parallel \nabla_\parallel f_{z1} + {\bs v}_{dz} \cdot \nabla f_{Mz},
\label{eq1storderdkefz}
\ee
where the independent velocity space coordinates are taken to be $\epsilon_z$ and $\mu_z$.
In a subsidiary expansion of equation~(\ref{eq1storderdkefz}) with respect to collisionality, the pressure anisotropy will appear in first order, as usual for a collisional species~\citep{braunhelander2010}. 
We define the collisionality here as $\nu_{*ab} = \nu_{ab}/\omega_{ta} = L_\parallel / \lambda_{mfp}^{ab}$, where $\omega_{ta}$ is the characteristic transit frequency of species $a$ along the magnetic field, $\nu_{ab}$ represents the characteristic collision frequency between species $a$ and $b$, and the mean free path $\lambda_{mfp}^{ab} = v_{Ta}/\nu_{ab}$. 
Comparing the collision and drift terms in eq.~(\ref{eq1storderdkefz}), remembering that the flows of all species are at the diamagnetic level $\sim \rho_{*a} v_{Ta}$, and that a factor $Ze\Phi_0/T$ is introduced through the gradient of $F_{Mz}$, we may expect $p_{z\parallel} - p_{z\perp} \sim Z p_z v_{dz}/\nu_{zz} L \sim Z p_z \rho_{*z} / \nu_{*zz}$. The parallel friction between unlike species drives the flux, and for the case of disparate mass ions considered here we may approximate it as $R_{zi\parallel} \sim m_i n_i \left(V_{i\parallel} - V_{z\parallel}\right) \nu_{iz} \sim m_i n_i \rho_{*i} v_{Ti} \nu_{iz}$, where $V_{i\parallel}$ and $V_{z\parallel}$ are the bulk ion and impurity parallel flows respectively. (The form of the collision operator is discussed in more detail in sections~\ref{secbulkcolop}-\ref{secimpdistrib}.)

We therefore find that the pressure anisotropy drive will be small when the collisionalities satisfy
\be
\frac{1}{\nu_{*iz}} \ll \frac{n_i}{n_z}\sqrt{\frac{m_i}{m_z}}\nu_{*zz}.
\label{eqcolconstraint}
\ee
When both species are collisional, as in~\citet{braunhelander2010}, this condition is clearly satisfied, even for non-trace impurity levels, and the pressure anisotropy drive is always small.
In the mixed collisionality case here, with $Z > 1$, this condition limits how collisionless the bulk ions can be compared to the impurities -- otherwise there would be a negligible frictional driving force.
We assume this ordering is satisfied and we will take the dominant drive of the transport in eq.~(\ref{eqgenradialflux}) to come from the parallel friction.
Momentum conservation in collisions then allows us to write the impurity flux in terms of the bulk ion-impurity parallel friction,
\be
R_{zi\parallel} = - R_{iz\parallel} = - m_i \int v_\parallel C_{iz}\left(f_{i},f_{z}\right)d^3v.
\label{eqdefnparfriction}
\ee
In the next subsections, we develop the expressions for the bulk ion and impurity distribution functions required to evaluate this friction, using model collision operators to treat the low collisionality regimes analytically.


\subsection{Bulk ion distribution function}
\label{secbulkdistrib}

The bulk ion distribution function can be treated throughout the low collisionality regimes of interest here using a recently developed formulation, which was detailed in~\citet{helanderetal2017b}. The distribution is split into pieces which are even and odd, $f^\pm_i$, with respect to the parallel velocity $v_\parallel = \sigma |v_\parallel|$, where $\sigma = \pm 1$. The full bulk ion drift kinetic equation then splits into two equations,
\be
v_\parallel \nabla_\parallel f_i^{\mp} = C_i^\pm (f_i) - {\bs v}_{di} \cdot \nabla f_i^{\pm},
\label{eqiondke}
\ee
where $C_i^\pm(f_i)$ denotes the even and odd parts of the collision operator $C_i(f_i)$ and the independent coordinates are taken to be $(r,\alpha,l,\epsilon_i,\mu_i,\sigma)$, where $\alpha$ labels different field lines on the same flux surface and $l$ gives the arc length along the magnetic field.

The orbit average may be introduced, which annihilates the left hand side of eq.~(\ref{eqiondke}) and is essentially a time average over the particle trajectory neglecting the drift motion. The parameter $\lambda = \mu_i / \epsilon_i$ divides phase space into regions describing particles trapped in the magnetic field structures, for which $\lambda > 1/B_{max}$ where $B_{max}(r)$ is the maximum value of the magnetic field strength on the flux surface, and those able to circulate freely. For circulating particles, the orbit average of an arbitrary function $g$ is defined as
\be
\overline{g}(r,\epsilon_i,\mu_i,\sigma) = \lim_{L\rightarrow \infty} \int_0^L g(r,\alpha,l,\epsilon_i,\mu_i)\frac{dl}{v_\parallel} \bigg\slash \int_0^L \frac{dl}{v_\parallel}.
\ee
This is independent of $\alpha$, as the integral extends along a field line so passes many times around the torus on a flux surface, and can also be written in terms of the flux surface average,
\be
\overline{g}(r,\epsilon_i,\mu_i,\sigma) = \lang \frac{Bg}{v_\parallel} \rang \bigg\slash \lang \frac{B}{v_\parallel} \rang.
\ee
In the trapped region the integral is taken between consecutive bounce points, denoted $l_1$ and $l_2$ at which $B(r,\alpha,l_1) = B(r,\alpha,l_2) = 1/\lambda$, so
\be
\overline{g}(r,\alpha,\epsilon_i,\mu_i) = \frac{1}{\tau_b} \int_{l_1}^{l_2} g^+(r,\alpha,l,\epsilon_i,\mu_i)\frac{dl}{|v_\parallel|},
\ee
where the bounce time $\tau_b = \int_{l_1}^{l_2} dl / |v_\parallel|$.

The odd piece of the distribution function is needed to evaluate the parallel friction in eq.~(\ref{eqdefnparfriction}). It was determined in~\citet{helanderetal2017a} for a pure plasma, where it was used to evaluate the parallel ion flow. For convenience we outline the arguments here, as we will finally evaluate different velocity space averages of the distribution and account for an impurity species.
Formally, the odd piece of the distribution follows from the line integral of the even eq.~(\ref{eqiondke}),
\be
f_i^-(r,\alpha,l,\epsilon_i,\mu_i,\sigma) = \int_{l_0}^l \left[C_i^+(f_i) - {\bs v}_{di} \cdot \nabla f_i^+\right]\frac{dl^\prime}{v_\parallel} + X\left(r,\alpha,\epsilon_i,\mu_i,\sigma\right).
\label{eqformalfodd}
\ee
We will return to the definition of $l_0$ momentarily. Ruling out the collisional limit for the bulk ions, the odd eq.~(\ref{eqiondke}) indicates that $f^+$ is a function of the constants of the motion, and the integration constant $X$ is then determined by the orbit average
\be
\overline{{\bs v}_{di} \cdot \nabla f_i^-} = \overline{C_i^-(f_i)}.
\label{eqoddconstraint}
\ee

The orbit average of the even equation~(\ref{eqiondke}) constrains the even piece of the distribution function appearing above,
\be
\overline{{\bs v}_{di} \cdot \nabla f_i^+} = \overline{C_i^+(f_i)}.
\label{eqevenconstraint}
\ee
This entails the assumptions on the quality of confinement noted in section~\ref{secintro}.
The ratio of the right to left-hand side of eq.~(\ref{eqevenconstraint}) is formally of the order $\nu_{*i}/\rho_{*i}$. In the $1/\nu$ regime, collisions are sufficiently dominant that the distribution function is nearly a Maxwellian and the derivation can proceed quite readily~\citep{helander2014}. At lower collisionality, orbit drifts can generate loss regions in velocity space, and the plasma is not generally in a local thermodynamic equilibrium. Two limits in which confinement can be adequately restored were described in~\citet{helanderetal2017a}. One is that in which the drift in the radial electric field, ${\bs v}_E = - \nabla \Phi \times {\bs b} / B$, is sufficiently strong compared to the magnetic drift, ${\bs v}_M$, that the bounce-averaged orbits stay close to a flux surface - this is often consistent with a large aspect ratio system. The other is when the orbit averaged magnetic drift is small compared to the local value, which is achieved when a stellarator is optimised to be near-omnigeneous. In both cases the distribution function is maintained near to Maxwellian, and is constant on a flux surface, thereby making the electrostatic potential a flux surface function, as assumed earlier.

We therefore assume here that either we are in the $1/\nu$ regime, or one of the above low collisionality conditions is satisfied. The even distribution can then be written in the form $f_i^+ = F_0 + F_1$, where $F_0(\epsilon_i,r)$ is a Maxwellian, and $F_1 \ll F_0$, remembering that it is constant along field lines, so is independent of $l$ in the trapped region of phase space, and independent of $\alpha$ and $l$ in the circulating region.
As the averaged drift $\overline{{\bs v}_{di} \cdot \nabla r} (\p_r F_0) = 0$ in the circulating region, it was argued in~\citet{helanderetal2017a} that $F_1$ is small in the circulating region, compared to its value in the trapped region, and we will neglect it. In the $1/\nu$ regime, $F_1 = 0$ also in the trapped region. In the lower collisionality regimes, the orbit average eq.~(\ref{eqevenconstraint}) requires $\overline{{\bs v}_d \cdot \nabla \alpha} (\p_\alpha F_1) + \overline{{\bs v}_d \cdot \nabla r} (\p_r F_0) \approx 0$. (The resolution of the behaviour of the distribution in the trapped-passing boundary layer is required to evaluate the bulk ion transport~\citep{hokulsrud1987}, but is not needed here.) The explicit drift term in eq.~(\ref{eqformalfodd}) can then be conveniently written for the low collisionality regimes of interest here as
\be
{\bs v}_{di} \cdot \nabla f_i^+ = \left({\bs v}_{di} \cdot \nabla r - \varepsilon_t \overline{{\bs v}_{di} \cdot \nabla r}\right)\frac{\p f_{Mi}}{\p r},
\label{eqvdfeven}
\ee
where $\varepsilon_t =0$ in the $1/\nu$ regime, and $\varepsilon_t =1$ in the trapped region of phase space and 0 otherwise in the $\sqrt{\nu}$ regime.

Now we consider the integration constant $X$. As the odd piece of the distribution function must vanish at a bounce point, if we choose $l_0$ in eq.~(\ref{eqformalfodd}) to be such a point, then $X=0$ in the trapped region. Therefore, we set:
\be
B(l_0) = \left\{
\begin{array}{cc}
1/\lambda & \lambda > 1/B_{max}, \\
B_{max} & \lambda < 1/B_{max}.
\end{array}
\right.
\ee
In the circulating region, $X$ is set by the constraint equation~(\ref{eqoddconstraint}). Using the conservative form of the particle drift, eq.~(\ref{eqconsvd}), along with the condition that circulating particles do not drift from their flux surfaces on average, it was shown in detail in~\citet{helanderetal2017a} that this constraint reduces to the following familiar form, for the low collisionality regimes of interest,
\be
\lang \frac{B}{v_\parallel} C_i^-(f_i) \rang = 0.
\label{eqpassconstraint}
\ee
In the next section we introduce a model collision operator which allows the integration constant to be determined explicitly, using this constraint.  We will then have, with eq.~(\ref{eqvdfeven}), the expression for $f_i^-$ needed to evaluate the moments giving the bulk ion flow and the impurity flux.


\subsection{Bulk ion collision operator}
\label{secbulkcolop}

The differences in the bulk ion flow in a pure plasma which result from using different forms of the collision operator to determine the odd piece of the distribution were discussed in~\citet{helanderetal2017a}. Similar considerations apply when evaluating particle fluxes via eqs.~(\ref{eqgenradialflux}) and~(\ref{eqdefnparfriction}). It is known that a momentum conserving collision operator is at least required to maintain the intrinsic ambipolarity of transport driven by friction. Therefore, we adopt here the following description of the bulk ion collisions.

Due to the disparate ion masses, we use a common approximation to the bulk ion-impurity collision operator $C_{iz}$~\citep{rosenbluthetal1972,helandersigmar2002},
\be
C_{iz}(f_i) = \nu_D^{iz}(v) \left( \mathcal{L}(f_{i}) + \frac{m_i v_\parallel V_{z\parallel}}{T} f_{Mi} \right).
\label{eqcizdefn}
\ee
The pitch angle scattering operator $\mathcal{L} = (1/2) \p_\xi \left( 1 - \xi^2 \right)\p_\xi$, where $\xi = \cos \theta = v_\parallel / v$ is the cosine of the particle pitch angle. With the normalised velocity $x_a = v/v_{Ta}$, the deflection frequency $\nu_D^{iz}(v) = 3\pi^{1/2}/4 \tau_{iz}x_i^3 = \hat{\nu}_D^{iz}/x_i^3$ and the collision time $\tau_{iz} = 3(2\pi)^{3/2}\sqrt{m_i}T^{3/2} \epsilon_0^2/n_{z} Z^2 e^4 \ln \Lambda$. The parallel impurity flow, $V_{z\parallel}$, will be determined in the next section.
Bulk ion self-collisions are described by an operator with a similar structure~\citep{rosenbluthetal1972,connor1973}, that is, a combination of pitch angle scattering and a momentum restoring term,
\be
C_{ii}(f_i) = \nu_D^{ii}(v) \left( \mathcal{L}\left(f_{i}\right) + \frac{m_i v_\parallel \mathcal{V}_{i\parallel}}{T} f_{Mi} \right).
\label{eqciidefn} 
\ee
The full energy dependent deflection frequency $\nu_D^{ii}(v) = \hat{\nu}_D^{ii}\left[\phi(x_i) - G(x_i)\right]/x_i^3$, $\hat{\nu}_D^{ii}$ is defined in analogy to $\hat{\nu}_D^{iz}$, the error function $\phi(x) = \left(2/\sqrt{\pi}\right)\int_0^x e^{-y^2} dy$ and the Chandrasekhar function $G(x) = \left[\phi(x) - x \phi^\prime(x)\right]/2x^2$.
The momentum restoring coefficient $\mathcal{V}_{i\parallel}$ will be set by requiring momentum conservation in bulk ion self-collisions, $\int v_\parallel C^-_{ii}\left(f_{i}\right) d^3v = 0$.
Altogether our model bulk ion collision operator is $C_i = C_{ii} + C_{iz}$, and we introduce the total collision frequency $\nu_D^i(v) = \nu_D^{ii} + \nu_D^{iz}$.
The bulk ion flow was evaluated in~\citet{helanderetal2017a} for the case $C_i = C_{ii}$, with $n_z = 0$, and as expected many similar steps appear in the derivation here. We highlight throughout the changes introduced by allowing for an impurity species.

For convenience we can set the electrostatic potential to zero on the surface of interest, and use the velocity space coordinate $\lambda = v_\perp^2 / v^2B$, which satisfies $\left. \nabla_\parallel \right|_{\epsilon,\mu} \lambda = 0$. The pitch angle scattering operator can be written as $\mathcal{L} = (2\xi/B) \p_\lambda \left(\lambda \xi \p_\lambda\right)$ and $\xi = \pm\sqrt{1-\lambda B}$.
The passing region constraint equation~(\ref{eqpassconstraint}) is then
\be
\frac{\p}{\p \lambda} \lambda \lang \sqrt{1-\lambda B} \frac{\p f_i^-}{\p \lambda} \rang + \frac{m_iv}{2T} \frac{\lang  \left(\nu_D^{ii} \mathcal{V}_{i\parallel} + \nu_D^{iz} V_{z\parallel}\right) B \rang}{\nu_D^i(v)} f_{Mi} =0.
\label{eqc0constraintmomcons}
\ee
Integrating over $\lambda$, with $\lambda < 1/B_{max}$, the integration constant vanishes upon requiring regularity at $\lambda = 0$.
We can now insert the general form for $f_i^-$ from eq.~(\ref{eqformalfodd}), noting that $F_1$ is taken to be negligible in the passing region, so the contribution from the term explicitly involving the collision operator vanishes.
We thus obtain a simple extension to eq.~(4.7) of~\citet{helanderetal2017a} to account for the presence of an impurity species,
\be
\frac{\p X}{\p \lambda}
= - \frac{m_iv}{2T_i \lang \xi\rang} \left( \frac{\lang \left(\nu_D^{ii} \mathcal{V}_{i\parallel} + \nu_D^{iz} V_{z\parallel}\right) B \rang}{\nu_D^i(v)}f_{Mi} + \frac{T}{e_i}\lang g_4 \rang \frac{\p f_{Mi}}{\p r}
\right),
\label{eqforconstc0momcons}
\ee
where the contribution from the drift term in eq.~(\ref{eqformalfodd}) gave rise to the known geometry function~\citep{nakajimaetal1989,helanderetal2011}
\be
g_4(\lambda,l) = \xi \int_{l_{max}}^l \left({\bf b} \times \nabla r \right)\cdot \nabla \xi^{-1} dl^\prime,
\label{eqdefng4}
\ee
with $\lambda < 1/B_{max}$ and $B(l_{max}) = B_{max}$.
The full form for the integration constant $X$ in the bulk ion distribution eq.~(\ref{eqformalfodd}) is thus given by eq.~(\ref{eqforconstc0momcons}), for $\lambda < 1/B_{max}$, and $X = 0$, for $1/B_{max} < \lambda < 1 /B_{min}$, where $B_{min}$ is the minimum field strength on the flux surface.

The momentum restoring coefficient, $\mathcal{V}_{i\parallel}$, is determined by momentum conservation in bulk ion self-collisions,
\be
0 = m_i \int v_\parallel C^-_{ii}(f_i) d^3v = m_i \int v_\parallel \nu_D^{ii} \left( \mathcal{L}(f_i^-) + \frac{m_iv_\parallel\mathcal{V}_{i\parallel}}{T} f_{Mi}\right) d^3v,
\ee
as the self-adjoint property of the Lorentz operator gives
\be
\mathcal{V}_{i\parallel} = \frac{1}{n_{i}  \left\{ \nu_D^{ii} \right\} } \int \nu_D^{ii} v_\parallel f_{i}^{-} d^3v.
\label{eqintforviparallel}
\ee
Here we have introduced the velocity space average~\citep{hirshman1976} for a function of the magnitude of the velocity, $\left\{ F(v) \right\} = (8/3\sqrt{\pi})\int_0^\infty F(x) x^4 e^{-x^2} dx$, so $ \left\{ \nu_D^{ii} \right\} \tau_{ii} = \sqrt{2} - \ln (1+\sqrt{2})$.
Inserting $f_i^-$ from eq.~(\ref{eqformalfodd}), we see as detailed in~\citet{helanderetal2017a} that the term explicitly containing $C^+(f_i)$ does not contribute when the collision operator is of the form assumed here. The explicit drift term is usefully written in terms of the function $u$ defined in section~(\ref{secfluxfrictiondef}), using the projection ${\bs v}_{di} \cdot \nabla r$ of eq.~(\ref{eqconsvd}), and results in the same contribution as in~\citet{helanderetal2017a}, with the integration constant in $u$ fixed by taking $u=0$ where $B=B_{max}$. The appearance of the impurity flow term in the integration constant here, however, gives an additional contribution compared to eq.~(4.12) of~\citet{helanderetal2017a},
\bea
& & - \frac{1}{n_i \left\{\nu_D^{ii}\right\}} \lang B^2 \int_0^\infty dv 2 \pi v^2 \nu_D^{ii} \int_0^{1/B_{max}} \lambda \frac{\p X}{\p \lambda} d\lambda \rang \nonumber \\
& & \qquad = \frac{f_c}{\left\{\nu_D^{ii}\right\}}\left( \left\{\frac{{\nu_D^{ii}}^2}{\nu_D^i}\right\} \lang B \mathcal{V}_{i\parallel} \rang + \left\{\frac{{\nu_D^{iz}}^2}{\nu_D^i}\right\}\lang BV_{z\parallel}\rang \right) + \frac{f_s T}{e}\left(A_{1i}-\eta A_{2i}\right),
\label{eqext412}
\eea
where it has been anticipated that we will only need the restoring coefficient in the form $\lang B \mathcal{V}_{i\parallel} \rang$. So we find  the following modification of eq.~(6.8) of~\citet{helanderetal2017a} in the presence of an impurity species,
\be
\lang B \mathcal{V}_{i\parallel} \rang \left[ 1 - \frac{f_c}{ \left\{ \nu_D^{ii} \right\}} \left\{\frac{{\nu_D^{ii}}^2}{\nu_D^i}\right\}\right] - \frac{f_c}{ \left\{ \nu_D^{ii} \right\}} \left\{\frac{\nu_D^{ii}\nu_D^{iz}}{\nu_D^i}\right\}\lang B V_{z\parallel} \rang= \frac{T}{e_i} \left(A_{1i} - \eta A_{2i} \right) \left[f_s + \lang (u+s)B^2\rang \right],
\label{eqconstraintforvi}
\ee
which reduces to that expression in the limit of a pure plasma, where $n_z \rightarrow 0$ and $\nu_D^i \rightarrow \nu_D^{ii}$.
Here $\eta =  \left\{ \nu_D^{ii}(5/2-x^2) \right\}/ \left\{ \nu_D^{ii} \right\} = (5/2) - 1/[2- \sqrt{2}\ln (1 + \sqrt{2})]$, 
\be
f_c  = \frac{3 \lang B^2 \rang}{4} \int_0^{1/B_{max}} \frac{\lambda d\lambda}{\lang \sqrt{1-\lambda B}\rang}, \qquad f_s  = \frac{3 \lang B^2 \rang}{4} \int_0^{1/B_{max}} \frac{\lang g_4\rang \lambda d\lambda}{\lang \sqrt{1-\lambda B}\rang},
\label{eqlambdaeffav}
\ee
and the term $s$ is zero in the $1/\nu$ regime, and given in the $\sqrt{\nu}$ regime by
\be
s(l) = \frac{3}{2}\int_{l_{max}}^l dl^\prime \int_{1/B_{max}}^{1/B(l^\prime)}\frac{d\lambda}{\xi(l^\prime)} \overline{\xi \left({\bs b} \times \nabla r \right) \cdot \nabla \left(\frac{\xi}{B}\right)}.
\ee

Finally, with the assumed quality of confinement described in section~\ref{secbulkdistrib} (that is $F_0 \approx f_{Mi}$) and the model collision operator, eq.~(\ref{eqcizdefn}), then adopted here, the parallel friction in eq.~(\ref{eqdefnparfriction}) needed to determine the particle flux takes the form
\be
R_{zi\parallel} = m_i \int \nu_D^{iz}(v) v_\parallel f_{i}^- d^3v - \frac{m_in_{i}}{\tau_{iz}} V_{z\parallel}.
\label{eqrizparmodelczi}
\ee
We therefore now need an expression for the parallel impurity flow and in the next section we consider the impurity distribution function.


\subsection{Impurity distribution function}
\label{secimpdistrib}

As introduced in section~(\ref{secfluxfrictiondef}), the collisional impurity species can be treated by the usual expansion of eq.~(\ref{eq1storderdkefz}) in the small parameter $1/\nu_{*zz}$~\citep{braunhelander2010}. At order $-1$, $C_z(f_{z1}^{(-1)}) =0$, so the impurity distribution has the form of a perturbed Maxwellian~\citep{helandersigmar2002},
\be
f_{z1}^{(-1)} = \left[\frac{p_z^{(-1)}}{p_{z}} + \frac{m_z}{T}v_\parallel V_{z\parallel}^{(-1)} + \left(x_z^2 - \frac{5}{2}\right) \frac{T_z^{(-1)}}{T}\right] f_{z0}.
\label{eqfzminus1all}
\ee
The parallel flow, $V_{z\parallel}^{(-1)}$, is constrained by momentum conservation in this order
\be
\int m_z v_\parallel C_{zi}\left(f_{z1}^{(-1)}\right) d^3v = 0,
\label{eqimpparflowconstraintorderminus1}
\ee
and we take the disparate mass form for the collision operator $C_{zi}$~\citep{hazeltinemeiss2003,helandersigmar2002},
\be
C_{zi}\left(f_{z1}\right) = - \frac{{\bs R}_{zi}}{m_zn_{z}}\cdot\frac{\p f_{z0}}{\p {\bs v}} + \frac{m_in_{i}}{m_zn_{z} \tau_{iz}}\frac{\p}{\p {\bs v}}\cdot \left[\left({\bs v} - {\bs V}_z\right) f_{z1} + \frac{T}{m_z}\frac{\p f_{z1}}{\p {\bs v}}\right].
\label{eqczidefn}
\ee
Using this in eq.~(\ref{eqimpparflowconstraintorderminus1}) gives simply $R_{zi\parallel}^{(-1)} = 0$.
Considering eq.~(\ref{eqrizparmodelczi}), this would require $V_{z\parallel}^{(-1)} \sim \rho_{*i} v_{Ti}$ here, that is $V_{z\parallel}^{(-1)}/v_{Tz} \sim \rho_{*z} Z$. However, by definition of the collisional expansion, $V_{z\parallel}^{(-1)} \sim \rho_{*z} v_{Tz} \nu_{*zz}$, that is $V_{z\parallel}^{(-1)}/v_{Tz} \sim \rho_{*z} Z ( \nu_{*iz} n_zZ / n_i)$. For the collisionless ions $ \nu_{*iz} \ll 1$, so restricting the impurity density such that $ \nu_{*iz} n_z Z / n_i < 1$ holds (hence we do not consider a pure ``impurity'' plasma) the two conditions give a contradiction.
This is resolved by requiring $V_{z\parallel}^{(-1)} = 0$, and so $R_{zi\parallel}^{(0)}$ is found to be the leading order friction driving the particle flux.

The form of the leading order flow, $V_{z\parallel} \approx V_{z\parallel}^{(0)}$, may also be found as usual by considering density conservation from the $v_\parallel/B$ moment of eq.~(\ref{eq1storderdkefz}) written in conservative form using eq.~(\ref{eqconsvd}), or radial force balance combined with incompressibility of the equilibrium flow in leading order:
\be
V_{z\parallel} = \left(\frac{1}{Ze n_z} \frac{dp_{z}}{dr} + \frac{d\Phi}{dr}\right)uB + \frac{K_z\left(r\right)B}{n_z}.
\label{eqgenformvpar}
\ee
A constraint on the flux surface function $K_z$ is obtained from the Spitzer-type problem for $f_{z1}^{(0)}$ arising at zeroth order in the collisional expansion of eq.~(\ref{eq1storderdkefz}),
\be
C_z\left(f_{z1}^{(0)}\right) = v_\parallel f_{z0} \left[A_{z1\parallel}^{(-1)} + \left(x_z^2 - \frac{5}{2}\right)A_{z2\parallel}^{(-1)}\right],
\label{eq0thorderimpdke}
\ee
where the parallel driving forces resulting from $f_{z1}^{(-1)}$ are
\be
A_{z1\parallel}^{(-1)} = \frac{\nabla_\parallel p_{z1}^{(-1)}}{p_z} + \frac{Ze}{T}\nabla_\parallel \phi_1^{(-1)}, \hspace{0.6cm} A_{z2\parallel}^{(-1)} = \frac{\nabla_\parallel T_{z1}^{(-1)}}{T}.
\ee
Parallel momentum conservation, that is the $m_z v_\parallel$ moment of eq.~(\ref{eq0thorderimpdke}), gives
\be
R_{zi\parallel}^{(0)} = n_{z}TA_{z1\parallel}^{(-1)}.
\label{eqparmomconsa1}
\ee
Upon taking the $B$-weighted flux surface average, the general property of the divergence of a vector field ${\bs F}$,
\be
\lang \nabla \cdot {\bs F} \rang = \frac{1}{V^\prime(r)}\frac{\p}{\p r} \lang V^\prime(r) {\bs F}\cdot \nabla r \rang,
\label{eqgenfsadivvector}
\ee
where $V$ is the volume enclosed by a flux surface, annihilates the parallel gradient terms and sets the constraint,
\be
\lang BR_{zi\parallel}^{(0)} \rang = 0.
\label{eqrzparconstraint}
\ee
This relation was first discussed in the context of transport in the mixed collisionality regime of a tokamak in~\citet{hirshman1976}.
Applying this to eq.~(\ref{eqrizparmodelczi}) results in
\be
\lang B V_{z\parallel} \rang = \frac{T}{Ze}A_{1z}  \lang uB^2 \rang + \frac{K_z(r)}{n_{z}} \lang B^2 \rang = \frac{\tau_{iz}}{n_{i}}\lang B \int \nu_D^{iz}(v) v_\parallel f_{i}^{-} d^3v \rang.
\label{eqconstraintforKz}
\ee
We will find in the following section that we do not need to solve explicitly for the function $K_z$ to determine the particle flux.


\section{Impurity flux}
\label{secflux}

With the ion distribution in eq.~(\ref{eqformalfodd}), and the constraint eqs.~(\ref{eqconstraintforvi}) and~(\ref{eqconstraintforKz}), we can now finalise the expression for the parallel friction driving the impurity flux in eq.~(\ref{eqgenradialflux}).
The integral needed in eq.~(\ref{eqrizparmodelczi}), and appearing in eq.~(\ref{eqconstraintforKz}), is very similar to that in the expression for the momentum restoring coefficient, eq.~(\ref{eqintforviparallel}), but with the simpler velocity dependence of $\nu_D^{iz}$, rather than $\nu_D^{ii}$.

Again the contribution from the collision operator vanishes and similar contributions arise from the explicit drift terms, resulting in
\be
R_{zi\parallel} = \frac{m_ip_i}{e\tau_{iz}}\left(A_{1i} - \frac{3}{2}A_{2i}\right)\left(u+s\right) B + P(r)B - \frac{m_in_{i}}{\tau_{iz}} V_{z\parallel},
\label{eqintrzipar}
\ee
where the flux function $P(r)$ contains the contribution resulting from the integration constant $X$,
\be
P(r) = \frac{m_ip_i}{e\tau_{iz}}\left[ \frac{f_s}{\lang B^2 \rang}\left(A_{1i} - \frac{3}{2}A_{2i}\right) + \frac{e \tau_{iz}}{T}\frac{f_c}{\lang B^2\rang}\left(\left\{\frac{\nu_D^{iz}\nu_D^{ii}}{\nu_D^i}\right\} \lang B \mathcal{V}_{i\parallel}\rang + \left\{\frac{{\nu_D^{iz}}^2}{\nu_D^i}\right\}\lang BV_{z\parallel}\rang\right)\right].
\label{eqforp}
\ee
Substituting for $V_{z\parallel}$ from eq.~(\ref{eqgenformvpar}) gives
\be
R_{zi\parallel} = \frac{m_ip_i}{e\tau_{iz}}\left[\left(A_{1i} - \frac{3}{2}A_{2i}\right)\left(u+s\right)B - \frac{A_{1z}}{Z}uB\right] + \left[P(r) - \frac{m_in_{i}}{n_z\tau_{iz}}K_z(r)\right]B,
\label{eqintrziparwithvz}
\ee
and we see that the friction has the following general structure
\be
R_{zi\parallel} = G_1(r) uB + G_2(r) sB + G_3(r) B, 
\label{eqrzing}
\ee
where the bulk ion momentum restoring coefficient and impurity flow coefficient $K_z$ only appear in the flux function $G_3$.
The impurity flow constraint gives
\be
\lang BR_{zi\parallel} \rang = G_1(r) \lang uB^2 \rang + G_2(r) \lang sB^2 \rang + G_3(r) \lang B^2 \rang = 0,
\label{eqg3constraint}
\ee
so we may eliminate $G_3$, and thus do not need to evaluate $\mathcal{V}_{i\parallel}$ or $K_z$ explicitly.
Finally then, the radial impurity flux is given by
\bea
\Gamma_z = \frac{1}{Ze}\lang uBR_{zi\parallel} \rang &=& - \frac{m_i p_i}{Ze^2 \tau_{iz}}\left[\frac{1}{Z}A_{1z}\left(\lang u^2 B^2 \rang  - \frac{\lang uB^2\rang^2}{\lang B^2 \rang}\right) \right. \label{eqimpflux} \\
& & - \left. \left(A_{1i} - \frac{3}{2}A_{2i}\right)\left(\lang u\left(u+s\right)B^2 \rang - \lang \left(u+s\right)B^2 \rang \frac{\lang uB^2 \rang}{\lang B^2 \rang}\right) \right]. \nonumber
\eea

The transport coefficients introduced in eq.~(\ref{eqdeffluxcoeffs}) can now be identified from the flux given in eq.~(\ref{eqimpflux}).
We can usefully note the appearance of the Pfirsch-Schl\"{u}ter coefficient in the flux of the collisional species, which can also be written in terms of the parallel current,
\be
 D_{PS} = \frac{m_iT_i}{e^2 \tau_{iz}} \left(\lang u^2B^2 \rang - \frac{\lang uB^2\rang^2}{\lang B^2 \rang}\right) = \frac{\rho_i^2}{\tau_{iz}}\frac{\lang J_\parallel^2\rang \lang B^2 \rang - \lang J_\parallel B\rang^2}{\left(dp/dr\right)^2},
\ee
and by the Schwartz inequality satisfies $D_{PS} \geq 0$. Therefore $D_{11}^{zz} = - n_i D_{PS} /Z^2 n_z$, and a given impurity density gradient drives an impurity flux in the opposite direction, as the increase of entropy requires. 
Note that the transport coefficients are independent of the impurity content, up to an overall density prefactor coming from $\tau_{iz}$.

When the bulk ions are in the $1/\nu$ regime, $s=0$ and $D_{11}^{zi}= -ZD_{11}^{zz}$, driving an impurity flux in the same direction as the bulk ion density gradient. The equality between the coefficients has also been shown to hold in the high collisionality limit, where both ion species are collisional~\citep{braunhelander2010}, and so the flux driven directly by the electric field {\it cancels out} in both of these regimes. We also see that $D_{12}^z = -(3/2)D_{11}^{zi}$, so there will be temperature screening when the bulk ions are in the $1/\nu$ regime, just as in a mixed collisionality tokamak~\citep{hirshman1976,samainwerkoff1977}. In the presence of a temperature gradient typically pointing inward, we thus expect an outward impurity flux to be driven if the logarithmic temperature gradient is more than twice that of the density, $\eta_i = \p \ln T_i / \p \ln n_i > 2$. This outward flux will not be overcome by any direct drive from the electric field, contrary to the expectation for lower-collisionality regimes. Note that such temperature screening is typically not the case in a collisional plasma~\citep{braunhelander2010,hirshman1977}, but an exception can occur in the very relevant case of a heavy impurity in a relatively clean plasma~\citep{rutherford1974,burrellwong1981}, where the effect of bulk ion friction dominates over that of impurity self-collisions.

As the bulk ions move into the lower collisionality $\sqrt{\nu}$ regime, the exact cancellation of the electric field drive coefficients is broken, leaving a drive which is proportional to the geometric quantity originating in the trapped particle drift, $\lang usB^2 \rang - \lang sB^2 \rang \lang uB^2 \rang / \lang B^2 \rang$. This is not sign definite and must be evaluated numerically for a given equilibrium, but we may expect it to be small in a well-optimised device.
The relation $D_{12}^z = -(3/2)D_{11}^{zi}$ remains valid, and depending on the sign of the geometric factor, either temperature screening will persist, or the bulk ion density gradient, typically pointing inward, will drive an additional outward impurity flux. The net flux, and strength of the drive by the electric field which typically points inward~\citep{hirschetal2008,klingeretal2017}, must finally be determined numerically in this low collisionality regime.

It is of interest to consider the tokamak limit of the above results, where $s=0$. The axisymmetric magnetic field can be written in the usual form: ${\bf B}= I(\psi) \nabla \phi + \nabla \phi \times \nabla \psi$, where $\psi$ the poloidal flux function is used as the radial coordinate, $\phi$ is toroidal angle and $I$ is related to the confining toroidal magnetic field~\citep{helandersigmar2002}, so the function $ u \rightarrow I \int_{l_{max}}^l\nabla_\parallel B^{-2} dl^\prime = I(B^{-2}-B_{max}^{-2})$. We then recover the well-known expression 
\be
\Gamma_z^{tok} = - \frac{m_i p_i I^2}{Ze^2 \tau_{iz}}\left(\frac{A_{1z}}{Z} - A_{1i} + \frac{3}{2}A_{2i}\right) \left( \lang \frac{1}{B^2}\rang - \frac{1}{\lang B^2\rang}\right),
\label{eqimpfluxtok}
\ee
first derived in~\citet{hirshman1976}, which shows temperature screening when the bulk ion temperature decreases radially, as expected.
We see that in the tokamak limit, the elimination of the function $G_3$ by eq.~(\ref{eqg3constraint}) represents the fact that the radial flux of a collisional species is driven only by the variation of the parallel friction on a flux surface.


\subsection{Two collisional impurities}
\label{sectwoz}

There are typically many impurity species present in magnetically confined fusion plasmas. A common situation is one in which there are trace amounts of a particularly heavy impurity, often released from the exhaust region, in a background of an otherwise dominant impurity, which may be released for example from the main walls. The transport of the heavier impurity is of particular importance, as it will be the most difficult to ionise and thus poses the strongest potential source of core radiation losses.
The results presented above allow us to make the following interesting observation when both impurity species are taken to be collisional, extending somewhat the analysis presented for the tokamak in~\citet{burrellwong1981}.

We denote the lighter impurity by a subscript $A$ here, with charge $Z_A \gg 1$, and continue to use $z$ for the heavier impurity. Following~\citet{braunhelander2010}, $V_{z\parallel}^{(-1)} = V_{A\parallel}^{(-1)} = 0$ as the species are collisional, and a flow cannot be driven at this order through interaction with the collisionless bulk.
Also, as species $A$ is collisional, the radial flux of species $z$ will continue to be dominated by the friction drive, as long as eq.~(\ref{eqcolconstraint}) is satisfied, so
\be
\Gamma_z  = \frac{1}{Ze} \lang uB R_{z\parallel} \rang = \frac{1}{Ze} \lang uB \left( R_{zi\parallel} + R_{zA\parallel} \right)\rang.
\label{eqgammaztwoimp}
\ee
Assuming that the bulk ions and species $A$ have disparate masses, $m_i \ll m_A$, collisions between them can be modelled by a collision operator analogous to that in eq.~(\ref{eqcizdefn}).
The contribution to the impurity flux from $R_{zi\parallel} = - R_{iz\parallel}$ can then be determined as a simple extension of the results above - we will again obtain eqs.~(\ref{eqintrzipar}) and~(\ref{eqintrziparwithvz}), but with the flux function $P(r)$ modified such that $\nu_D^i \rightarrow \nu_D^{ii} + \nu_D^{iz} + \nu_D^{iA}$ and
\be
\left\{\frac{{\nu_D^{iz}}^2}{\nu_D^i} \right\} \lang BV_{z\parallel} \rang \rightarrow \left\{\frac{{\nu_D^{iz}}^2}{\nu_D^i} \right\} \lang BV_{z\parallel} \rang + \left\{\frac{\nu_D^{iz} \nu_D^{iA}}{\nu_D^i} \right\} \lang BV_{A\parallel} \rang.
\ee
Parallel momentum constraints analogous to eq.~(\ref{eqrzparconstraint}) are obtained similarly for the two collisional species,
\bea
\lang B\left(R_{zi\parallel} + R_{zA\parallel} \right)\rang = 0 \nonumber \\
\lang B\left(R_{Ai\parallel} + R_{Az\parallel} \right)\rang = 0.
\label{eqtwoimpconstraints}
\eea
The first of these allows us to again eliminate $G_3$ from eq.~(\ref{eqrzing}) leaving
\be
R_{zi\parallel} = G_1(r) \left(uB - \lang uB^2 \rang \frac{B}{\lang B^2 \rang} \right) + G_2(r) \left(sB - \lang sB^2 \rang \frac{B}{\lang B^2 \rang} \right) - \frac{B}{\lang B^2\rang}\lang BR_{zA\parallel} \rang.
\label{eqrzipartwoimp}
\ee

Note that the total radial impurity current is
\be
J_{imp} \equiv e_z \Gamma_z + e_A \Gamma_A = - \lang uB \left(R_{iz\parallel} + R_{iA\parallel}\right)\rang.
\ee
The disparate mass collision operator adopted will lead to an expression for $R_{iA\parallel} = -R_{Ai\parallel}$ analogous to eq.~(\ref{eqintrziparwithvz}). Summing the two constraints in eq.~(\ref{eqtwoimpconstraints}) gives $\lang B\left(R_{iz\parallel} + R_{iA\parallel} \right)\rang = 0$, which allows all of the unknown flux functions $\lang B\mathcal{V}_{i\parallel} \rang$, $K_A$ and $K_z$ to again be eliminated from the flux, leaving
\bea
J_{imp} &=& - \frac{m_i p_i}{e \tau_{iz}}\left[\left(\frac{A_{1z}}{Z} + \frac{1}{\zeta_A}\frac{A_{1A}}{Z_A}\right)\left(\lang u^2 B^2 \rang  - \frac{\lang uB^2\rang^2}{\lang B^2 \rang}\right) \right. \\
& & - \left. \left(1+\frac{1}{\zeta_A}\right)\left(A_{1i} - \frac{3}{2}A_{2i}\right)\left(\lang u\left(u+s\right)B^2 \rang - \lang \left(u+s\right)B^2 \rang \frac{\lang uB^2 \rang}{\lang B^2 \rang}\right) \right], \nonumber
\eea
where $\zeta_A = n_z Z^2 / n_A Z_A^2$. The total impurity current can thus also experience temperature screening, under the conditions described in the previous section.

To form the explicit expression for the flux of the heavier impurity, we still need to determine the combination
\be
\lang uB R_{zA\parallel} \rang - \frac{\lang uB^2\rang}{\lang B^2 \rang} \lang BR_{zA\parallel} \rang,
\label{eqrzacontrib}
\ee
where the friction $R_{zA\parallel} = m_z \int v_\parallel C_{zA}(f_z,f_A) d^3v$ contains the linearised collision operator $C_{zA}$ acting on the distribution functions of the two collisional species. These are given to leading order by the solution of eq.~(\ref{eq0thorderimpdke}) and the analogous equation for $f_{A1}^{(0)}$. The solution can be written as an expansion in Sonine polynomials, $L_\alpha^{(3/2)}(x^2)$, such that
\be
f_{a1}^{(0)} = \sum_{\alpha=0}^2 u_{a\alpha} L_\alpha^{(3/2)}(x_a^2) \frac{m_av_\parallel}{T} f_{Ma}.
\label{eqcolfexp}
\ee
With $L_0^{(3/2)}(x^2) = 1$ and $L_1^{(3/2)}(x^2) = (5/2)-x^2$, the expansion coefficients may be recognised as $u_{a0} = V_{a\parallel}$ and $u_{a1} = -2q_{a\parallel}/5p_a$, where $q_{a\parallel}$ is the parallel heat flux.
Substituting this expansion into $R_{zA\parallel}$, the integration over the collision operator may be performed directly~\citep{helandersigmar2002}, and the parallel friction coefficients will depend on the mass ratio of the impurities. We treat the case of disparate impurity masses, $m_A \ll m_z$ and $Z_A \ll Z$, explicitly here, which may give a good approximation to the experimentally relevant case of a low collisionality bulk H plasma with a main impurity such as C from the walls, and a low density, heavier component, such as Fe. Then
\be
R_{zA\parallel} = \frac{m_i n_i}{\tau_{iz}}\sqrt{\frac{m_A}{m_i}}\frac{n_AZ_A^2}{n_i}\left( V_{A\parallel} - V_{z\parallel} - \frac{3}{5} \frac{q_{A\parallel}}{p_A} + \frac{15}{8}u_{A2} \right).
\label{eqrziparcoeffs}
\ee
The $v_\parallel L_2$-moment of eq.~(\ref{eq0thorderimpdke}) for species $A$ relates the coefficient $u_{A2}$ to the parallel flows in the disparate mass limit,
\be
u_{A2} = \frac{1}{\left(45\sqrt{2} + 433\zeta_A/4\right)}\left[30 \zeta_A \left(V_{z\parallel} - V_{A\parallel}\right) + \frac{2}{5}\left(12 \sqrt{2} + 69 \zeta_A\right)\frac{q_{A\parallel}}{p_A}\right].
\label{equa2}
\ee

The parallel species flows in eq.~(\ref{eqrziparcoeffs}), $V_{z\parallel}$ and $V_{A\parallel}$, have the general form of eq.~(\ref{eqgenformvpar}). In the combination of eq.~(\ref{eqrzacontrib}), all terms containing the flux functions $K_z$ and $K_A$ cancel, leaving only contributions to the impurity flux from the radial gradients $A_{1z}$ and $A_{1A}$. The form of the parallel heat flows in eq.~(\ref{eqrziparcoeffs}) can be determined using the $v_\parallel \epsilon_z/B$ moment of the conservative form of eq.~(\ref{eq1storderdkefz}) (and the analogous equations for species $A$ and $i$), which gives the equation of energy conservation for a species,
\be
B \nabla_\parallel \left(\frac{q_{a\parallel}}{B} - \frac{5}{2}\frac{p_a T}{e_a}A_{2a}u\right) = \int \frac{m_a v^2}{2}C_a(f_{a1}) d^3v.
\label{eqgenenergycons}
\ee
The energy exchange between species appearing on the right hand side competes with the parallel heat flux to determine the parallel temperature perturbation on a flux surface. It is typified here for disparate mass species using the second term of eq.~(\ref{eqczidefn}), giving for example
\be
\int \frac{m_z v^2}{2}C_{zi}(f_{z})d^3v = \frac{3 m_i n_i}{m_z \tau_{iz}}\left(T_i - T_z\right),
\ee
between the heaviest impurity and the bulk ions (remember to leading order here the ion temperatures are equal, which leaves only the perturbed temperatures in this expression).
The $v_\parallel L_1$ and $v_\parallel L_2$-moments of eq.~(\ref{eq0thorderimpdke}) give us, in the disparate mass case, the heavy impurity parallel heat flux
\be
q_{z\parallel} = - \frac{125 \sqrt{2}}{32}\frac{p_z^2 \tau_{zz}}{n_z m_z} A_{z2\parallel}^{(-1)}.
\ee
Thus we see that energy exchange with the low collisionality bulk ions can only be neglected when $1 \gg (n_i / n_z)\sqrt{m_i/m_z} \nu_{*zz} \nu_{*iz}$, which cannot be satisfied consistently with the condition eq.~(\ref{eqcolconstraint}). This arises similarly for energy exchange between the impurity species $A$ and the bulk ions.
Energy exchange between the collisional impurity species is dominant when $1 \ll (Z^2/Z_A^2)\sqrt{m_A/m_z} \nu_{*zz} \nu_{*AA}$, which will always be satisfied.
Therefore we take the perturbed temperature of each impurity species to be equal, and set by energy exchange to that of the collisionless bulk ions. The parallel temperature gradients will then be negligible, and so the parallel impurity heat fluxes can be neglected in the expressions above.

The flux of the heaviest impurity can now be constructed from eqs.~(\ref{eqgammaztwoimp}),~(\ref{eqrzipartwoimp}), and~(\ref{eqrziparcoeffs}), with eq.~(\ref{equa2}) and the parallel flows just discussed, giving the final form
\bea
\lang \Gamma_z \cdot \nabla r \rang &=& - \frac{m_i p_i}{Ze^2 \tau_{iz}}\left\{\left[\left(1 + \sqrt{\frac{m_A}{m_i}}\frac{n_A Z_A^2}{n_i} \left(1 - Y\right) \right)\frac{A_{1z}}{Z} - \sqrt{\frac{m_A}{m_i}}\frac{n_A Z_A^2}{n_i}\left(1 - Y\right)\frac{A_{1A}}{Z_A}\right] \right. \nonumber \\
& & \hspace{2cm} \times \left(\lang u^2 B^2 \rang  - \frac{\lang uB^2\rang^2}{\lang B^2 \rang}\right) \nonumber \\
& & - \left. \left(A_{1i} - \frac{3}{2}A_{2i}\right)\left(\lang u\left(u+s\right)B^2 \rang - \lang \left(u+s\right)B^2 \rang \frac{\lang uB^2 \rang}{\lang B^2 \rang}\right) \right\},
\eea
where $Y = 225\zeta_A / (180\sqrt{2} + 433\zeta_A)$.
The net drive from the electric field still vanishes in the $1/\nu$ ($s=0$) regime.
The second impurity enhances the flux driven by the impurity density gradient, whilst introducing an oppositely directed component to the flux, when both impurity density gradients have the same sign. The net effect of introducing a second collisional species thus depends on the combination $(Z^{-1}A_{1z} - Z_A^{-1}A_{1A})$, producing an additional outward contribution to the flux when this quantity is negative.
Note that the result above does not require that the heaviest impurity $z$ is only present in trace quantities, but does also correctly describe that case.


\section{Bulk ion flow}
\label{secflow}

In this section we determine the bulk ion flow parallel to the magnetic field in a mixed collisionality plasma, returning to the case where only a single collisional impurity species is present. The flow is needed to evaluate the bootstrap current, which was considered for a pure plasma in the low collisionality $1/\nu$ and $\sqrt{\nu}$ regimes in~\citet{helanderetal2017a}.
The bulk ion parallel flow has the same general form as that of the impurities in eq.~(\ref{eqgenformvpar}), and it is in order to determine the equivalent flux surface function, $K_i(r)$, that we require a kinetic solution. 

We must evaluate the integral of the bulk ion distribution function,
\be
\lang V_{i\parallel} B \rang = \frac{1}{n_i} \lang B \int v_\parallel f_{i}^{-} d^3v \rang.
\label{eqformalvparb}
\ee
As discussed in section~\ref{secbulkcolop}, such an integral was considered in~\citet{helanderetal2017a} with the momentum conserving bulk ion self-collision operator used here. The similar structure of the disparate mass bulk ion-impurity collision operator adopted here allows the form of the integral to be given readily upon inserting the odd piece of the distribution, eq.~(\ref{eqformalfodd}), into eq.~(\ref{eqformalvparb}) and following the procedure of~\citet{helanderetal2017a}. The term containing the collision operator is seen to vanish due to particle conservation in collisions, while the term containing the drifts recovers eq.~(6.4) of~\citet{helanderetal2017a}. 
The effect of the additional impurity collisions again appears through their contribution to the integration constant, extending eq.~(4.17) of~\citet{helanderetal2017a} analogously to eq.~(\ref{eqext412}) here. This produces the modified flow expression
\be
\lang V_{i\parallel} B \rang = \frac{T}{e} A_{1i}\left(f_s + \lang\left(u+s\right)B^2\rang  \right) + f_c \lang B\mathcal{V}_{i\parallel} \rang \left(\left\{ \frac{\nu_D^{ii}}{\nu_D^i}\right\}  + \left\{ \frac{\nu_D^{iz}}{\nu_D^i}\right\} \frac{\lang BV_{z\parallel} \rang}{\lang B\mathcal{V}_{i\parallel} \rang} \right).
\label{eqviparbfsaflowform}
\ee
We can finally eliminate the inter-dependent flux surface averaged quantities $\lang B\mathcal{V}_{i\parallel} \rang$ and $\lang BV_{z\parallel} \rang$ appearing here. The expression for $\lang B\mathcal{V}_{i\parallel} \rang$ was given in eq.~(\ref{eqconstraintforvi}) and the integral on the right hand side of eq.~(\ref{eqconstraintforKz}) giving $\lang BV_{z\parallel} \rang$ was evaluated in section~\ref{secflux}, leading to the first two terms on the right hand side of eq.~(\ref{eqintrzipar}).
Thus we can form the ratio
\be
\frac{\lang BV_{z\parallel} \rang}{\lang B\mathcal{V}_{i\parallel} \rang} = \frac{\left(A_{1i} - \frac{3}{2}A_{2i}\right)\left(1 - \frac{f_c}{\left\{\nu_D^{ii}\right\}} \left\{\frac{{\nu_D^{ii}}^2}{\nu_D^i}\right\} \right) + \left(A_{1i} - \eta A_{2i}\right)f_c \tau_{iz} \left\{\frac{\nu_D^{iz}\nu_D^{ii}}{\nu_D^i}\right\} }{\left(A_{1i} - \eta A_{2i}\right)\left(1 - f_c \tau_{iz} \left\{\frac{{\nu_D^{iz}}^2}{\nu_D^i}\right\} \right) + \left(A_{1i} - \frac{3}{2}A_{2i}\right)\frac{f_c}{\left\{\nu_D^{ii}\right\}} \left\{\frac{\nu_D^{ii}\nu_D^{iz}}{\nu_D^i}\right\} },
\ee
and extract
\be
\lang B\mathcal{V}_{i\parallel} \rang = \frac{T}{e} \left(f_s + \lang (u+s)B^2\rang \right)  \frac{\left(A_{1i} - \eta A_{2i}\right)\left(1 - f_c \tau_{iz} \left\{\frac{{\nu_D^{iz}}^2}{\nu_D^i}\right\} \right) + \left(A_{1i} - \frac{3}{2}A_{2i}\right)\frac{f_c}{\left\{\nu_D^{ii}\right\}} \left\{\frac{\nu_D^{ii}\nu_D^{iz}}{\nu_D^i}\right\} }{\left(1 - \frac{f_c}{\left\{\nu_D^{ii}\right\}} \left\{\frac{{\nu_D^{ii}}^2}{\nu_D^i}\right\} \right)\left(1 - f_c \tau_{iz} \left\{\frac{{\nu_D^{iz}}^2}{\nu_D^i}\right\} \right) - \frac{f_c^2 \tau_{iz}}{\left\{\nu_D^{ii}\right\}} \left\{\frac{\nu_D^{ii}\nu_D^{iz}}{\nu_D^i}\right\}^2}.
\label{eqvpariconstfull}
\ee

The bulk ion contribution to the bootstrap current can be written in terms of transport coefficients as follows
\be
\lang J_{i\parallel} B \rang = n_i e \lang V_{i\parallel} B \rang= p_{i} \left(\mathcal{L}_{31}^{ii}A_{1i} + \mathcal{L}_{32}^{ii}A_{2i}\right).
\ee
These coefficients can be identified directly from eqs.~(\ref{eqviparbfsaflowform}-\ref{eqvpariconstfull}). However, to clarify the expressions analytically, we now assume a simplified dependence of the bulk ion self-collision frequency on velocity~\citep{newtonhelander2006}, taking it to have the same form as the bulk ion-impurity collision frequency introduced in eq.~(\ref{eqcizdefn}). This gives
\be
\frac{\nu_D^{iz}}{\nu_D^{ii}} \approx \frac{\tau_{ii}}{\tau_{iz}} = \frac{n_z Z_z^2}{n_i}  \equiv \zeta,
\label{eqapproxfornuiicolfreq}
\ee
where the parameter $\zeta$ usefully represents the impurity content. Defining the effective charge $Z_{\rm eff} = \sum_{a=i,z} n_a Z_a^2 / n_e$, the approximation in eq.~(\ref{eqapproxfornuiicolfreq}) reproduces the correct limits for $Z_{\rm eff} \rightarrow 1$ and $Z_{\rm eff} \rightarrow \infty$, and when in the trace limit, $n_z Z \ll n_i$, reduces to the familiar $Z_{eff} \approx 1 + \zeta$. The last term of eq.~(\ref{eqviparbfsaflowform}) now simplifies to
\be
\left(\left\{\frac{\nu_D^{ii}}{\nu_D^i}\right\} + \left\{\frac{\nu_D^{iz}}{\nu_D^i}\right\} \frac{\lang BV_{z\parallel}\rang}{\lang B\mathcal{V}_{i\parallel} \rang}\right) \lang B\mathcal{V}_{i\parallel}\rang = \frac{T}{e}\frac{\left(f_s + \lang (u+s)B^2\rang \right) }{1-f_c} \left[A_{i1} - \frac{3}{2}\frac{\left(\zeta + 2\eta/3\right)}{1+\zeta} A_{2i}\right].
\ee
Thus we have a generalisation of eq.~(6.9) of~\citet{helanderetal2017a} to the case of a mixed collisionality plasma with finite impurity content,
\be
\lang J_{i\parallel} B \rang =  p_{i}\frac{\left(f_s + \lang (u+s)B^2\rang \right) }{1 - f_c} \left[A_{1i} - \frac{3}{2}f_c\frac{\left(\zeta + 2\eta/3\right)}{1+\zeta}A_{2i}\right].
\label{eqjionapprox}
\ee
Note that, when $s=0$, the contribution from the radial electric field is cancelled by a similar contribution to the electron bootstrap current~\citep{helanderetal2017a}, making the total current independent of $E_r$ in the $1/\nu$ regime. 

We can see from eq.~(\ref{eqviparbfsaflowform}) that if bulk ion collisions are approximated by pure pitch angle scattering (PAS) the momentum restoring terms do not appear, so $\lang J_{i\parallel} B \rang^{PAS} = p_{i} A_{1i} \left(f_s + \lang (u+s)B^2\rang \right)$ and the effect of the impurities only enters through the alteration of the main ion density in the prefactor. Accounting for momentum conservation in collisions introduces $\mathcal{L}_{32}^{ii}$, which has an explicit dependence on impurity content.
In the axisymmetric tokamak limit, with $s=0$, $f_s \rightarrow I(f_c - \lang B^2\rang/B_{max}^2)$ and $f_s - \lang uB^2\rang \rightarrow -I(1-f_c)$, and so we recover the expression for the bulk ion current in the presence of impurities~\citep{newtonhelander2006,fieldetal2009}.


\section{Discussion}
\label{secdiscussion}

Neoclassical impurity accumulation in the core of stellarator plasmas, under the action of the radial electric field, has long been considered inevitable.
The conclusion was based on simplified models of the collisional interaction between species.
We have extended the treatment of stellarator impurity transport to the mixed collisionality regime, using a general flux-friction relation which was introduced previously to treat collisional plasmas. In this experimentally relevant regime, a heavy, highly charged, collisional impurity is taken to be present in a hydrogenic, bulk plasma, with the bulk ions in one of the low collisionality stellarator regimes. Here we have treated specifically the $1/\nu$ and $\sqrt{\nu}$ regimes, assuming the electric field is sufficiently strong or the geometry is sufficiently well optimised that the plasma is well confined.
The impurity flux is then dominated by the drive from friction against the bulk ions, with the formal requirement set by eq.~(\ref{eqcolconstraint}).

The results here show that in the mixed collisionality limit, impurity temperature screening will occur when the bulk ions are in the $1/\nu$ regime, if the logarithmic temperature gradient is more than twice the logarithmic density gradient, $\eta_i >2$.
In the appropriate limit, the impurity flux reduces to that of a tokamak, where such a screening effect is expected.
Furthermore, the direct drive of the impurity flux by the electric field vanishes, contrary to the usual expectation, when the bulk ions are in the $1/\nu$ regime. This feature does not hold as the bulk ions move into the lower collisionality $\sqrt{\nu}$ regime, as an additional geometric factor appears in the bulk ion gradient drive terms, originating in the orbit average of the trapped particle drift. This factor may be expected to be small in a well-optimised stellarator, which would result in an impurity flux driven only weakly by the electric field, and a weakly affected temperature screening.
As the proportionality between the bulk ion density and temperature gradient drives is maintained throughout the two low collisonality regimes considered here, any reduction in temperature screening is accompanied by an increased outward flux of impurities driven by the bulk ion density gradient.
The net direction of the remaining, small impurity flux will thus have to be determined numerically in the lower collisionality regime. In practice, this flux is so small that it may be overwhelmed by turbulent transport.

The presence of a second, lighter, collisional impurity species is found to enhance the flux of the heaviest impurity driven by its own density gradient. However, it also introduces a flux driven in the opposite direction, by the density gradient of the second species, which may be expected to dominate and give a typically inward contribution to the flux.

We will present a numerical study of the transport coefficients derived here in an upcoming paper, using the neoclassical code SFINCS~\citep{landremanetal2014}. This is a continuum $\delta f$ code, which can treat multiple species with the full linearised Landau collision operator. A summary of the initial successful comparison was given in~\citet{helanderetal2017b}.  
Note that numerical indications of temperature screening were already seen in~\citet{mollenetal2015}, and the analysis presented here and summarised in~\citet{helanderetal2017b} provides an explanation of those results.

Finally, the calculation of the radial flux by a flux-friction relation here used the piece of the bulk ion distribution which is odd in the parallel velocity. With this we could also evaluate the bulk ion contribution to the bootstrap current, which must be well-controlled in a stellarator with an island divertor, such as W7-X, and consider the effect of an impurity species. We see as usual that the inclusion of momentum restoring terms in the collision operator can introduce a substantial change to the expected flow, and strongly modify the dependence on impurity content.


We thank Craig Beidler, Felix Parra, Matt Landreman, Istvan Pusztai, John Omotani and T\"{u}nde F\"{u}l\"{o}p for helpful discussions, and acknowledge the hospitality of Merton College, Oxford, where this work was initiated.
This work was supported by the Framework grant for Strategic Energy
Research (Dnr. 2014-5392) from Vetenskapsr{\aa}det.

%

\bibliographystyle{jpp}

\bibliography{MixedColJPPpaper}

\begin{thebibliography}{35}
\expandafter\ifx\csname natexlab\endcsname\relax\def\natexlab#1{#1}\fi
\def\au#1{#1} \def\ed#1{#1} \def\yr#1{#1}\def\at#1{#1}\def\jt#1{\textit{#1}}
  \def\bt#1{#1}\def\bvol#1{\textbf{#1}} \def\vol#1{#1} \def\pg#1{#1}
  \def\publ#1{#1}\def\arxiv#1{#1}\def\org#1{#1}\def\st#1{\textit{#1}}

\bibitem[Angioni \& Helander(2014)]{angionihelander2014}
{\sc \au{Angioni, C.} \& \au{Helander, P.}} \yr{2014}  \at{Neoclassical
  transport of heavy impurities with poloidally asymmetric density distribution
  in tokamaks}.  \jt{Plasma Phys. Control. Fusion}  \bvol{56},  \pg{124001}.

\bibitem[Beidler {\em et~al.\/}(2011)Beidler, Allmaier, Isaev, Kasilov,
  Kernbichler, Leitold, Maa{\ss}berg, Mikkelsen, Murakami, Schmidt, Spong,
  Tribaldos \& Wakasa]{beidleretal2011}
{\sc \au{Beidler, C.~D.}, \au{Allmaier, K.}, \au{Isaev, M.~Y.}, \au{Kasilov,
  S.~V.}, \au{Kernbichler, W.}, \au{Leitold, G.~O.}, \au{Maa{\ss}berg, H.},
  \au{Mikkelsen, D.~R.}, \au{Murakami, S.}, \au{Schmidt, M.}, \au{Spong,
  D.~A.}, \au{Tribaldos, V.} \& \au{Wakasa, A.}} \yr{2011}  \at{Benchmarking of
  the mono-energetic transport coefficients -- results from the {International
  Collaboration on Neoclassical Transport in Stellarators (ICNTS)}}.  \jt{Nucl.
  Fusion}  \bvol{51},  \pg{076001}.

\bibitem[Braun \& Helander(2010)]{braunhelander2010}
{\sc \au{Braun, S.} \& \au{Helander, P.}} \yr{2010}  \at{{Pfirsch-Schl\"{u}ter}
  impurity transport in stellarators}.  \jt{Phys. Plasmas}  \bvol{17},
  \pg{072514}.

\bibitem[Burrell \& Wong(1981)]{burrellwong1981}
{\sc \au{Burrell, K.~H.} \& \au{Wong, S.~K.}} \yr{1981}  \at{Transport of a
  trace impurity in a dirty plasma in the {Pfirsch-Schl\"{u}ter} regime}.
  \jt{Phys. Fluids}  \bvol{24},  \pg{284--289}.

\bibitem[Connor(1973)]{connor1973}
{\sc \au{Connor, J.~W.}} \yr{1973}  \at{The neo-classical transport theory of a
  plasma with multiple ion species}.  \jt{Plasma Phys.}  \bvol{15},
  \pg{765--782}.

\bibitem[Field {\em et~al.\/}(2009)Field, McCone, Conway, Dunstan, Newton \&
  Wisse]{fieldetal2009}
{\sc \au{Field, A.~R.}, \au{McCone, J.}, \au{Conway, N.~J.}, \au{Dunstan, M.},
  \au{Newton, S.} \& \au{Wisse, M.}} \yr{2009}  \at{Comparison of measured
  poloidal rotation in {MAST} spherical tokamak plasmas with neo-classical
  predictions}.  \jt{Plasma Phys. Control. Fusion}  \bvol{51},  \pg{105002}.

\bibitem[{Garc\'{i}a-Rega\~{n}a} {\em et~al.\/}(2017){Garc\'{i}a-Rega\~{n}a},
  Beidler, Kleiber, Helander, Moll\'{e}n, Alonso, Landreman, Maa{\ss}berg,
  Smith, Turkin \& Velasco]{garciareganaetal2017}
{\sc \au{{Garc\'{i}a-Rega\~{n}a}, J.~M.}, \au{Beidler, C.~D.}, \au{Kleiber,
  R.}, \au{Helander, P.}, \au{Moll\'{e}n, A.}, \au{Alonso, J.~A.},
  \au{Landreman, M.}, \au{Maa{\ss}berg, H.}, \au{Smith, H.~M.}, \au{Turkin, Y.}
  \& \au{Velasco, J.~L.}} \yr{2017}  \at{Electrostatic potential variation on
  the flux surface and its impact on impurity transport}.  \jt{Nucl. Fusion}
  \bvol{57},  \pg{056004}.

\bibitem[Geiger {\em et~al.\/}(2015)Geiger, Beidler, Feng, Maa{\ss}berg,
  Marushchenko \& Turkin]{geigeretal2015}
{\sc \au{Geiger, J.}, \au{Beidler, C.~D.}, \au{Feng, Y.}, \au{Maa{\ss}berg,
  H.}, \au{Marushchenko, N.~B.} \& \au{Turkin, Y.}} \yr{2015}  \at{Physics in
  the magnetic configuration space of {W7-X}}.  \jt{Plasma Phys. Control.
  Fusion}  \bvol{57},  \pg{014004}.

\bibitem[Hazeltine \& Meiss(2003)]{hazeltinemeiss2003}
{\sc \au{Hazeltine, R.~D.} \& \au{Meiss, J.~D.}} \yr{2003} {\em Plasma
  Confinement\/}.  \publ{Dover Publications}.

\bibitem[Helander(2014)]{helander2014}
{\sc \au{Helander, P.}} \yr{2014}  \at{Theory of plasma confinement in
  non-axisymmetric magnetic fields}.  \jt{Rep. Prog. Phys.}  \bvol{77},
  \pg{087001}.

\bibitem[Helander {\em et~al.\/}(2012)Helander, D.Beidler, Bird, Drevlak, Feng,
  Hatzky, Jenko, Kleiber, Proll, Turkin \& Xanthopoulos]{helanderetal2012}
{\sc \au{Helander, P.}, \au{D.Beidler, C.}, \au{Bird, T.~M.}, \au{Drevlak, M.},
  \au{Feng, Y.}, \au{Hatzky, R.}, \au{Jenko, F.}, \au{Kleiber, R.}, \au{Proll,
  J. H.~E.}, \au{Turkin, Y.} \& \au{Xanthopoulos, P.}} \yr{2012}
  \at{Stellarator and tokamak plasmas: a comparison}.  \jt{Plasma Phys. and
  Control. Fusion}  \bvol{54},  \pg{124009}.

\bibitem[Helander {\em et~al.\/}(2011)Helander, Geiger \&
  Maa{\ss}berg]{helanderetal2011}
{\sc \au{Helander, P.}, \au{Geiger, J.} \& \au{Maa{\ss}berg, H.}} \yr{2011}
  \at{On the bootstrap current in stellarators and tokamaks}.  \jt{Phys.
  Plasmas}  \bvol{18},  \pg{092505}.

\bibitem[Helander {\em et~al.\/}(2017{\natexlab{{\em a\/}}})Helander, Newton,
  Moll\'{e}n \& Smith]{helanderetal2017b}
{\sc \au{Helander, P.}, \au{Newton, S.~L.}, \au{Moll\'{e}n, A.} \& \au{Smith,
  H.~M.}} \yr{2017{\natexlab{{\em a\/}}}}  \at{Impurity transport in a mixed
  collisionality stellarator plasma}.  \jt{Phys. Rev. Lett.}  \bvol{118},
  \pg{155002}.

\bibitem[Helander {\em et~al.\/}(2017{\natexlab{{\em b\/}}})Helander, Parra \&
  Newton]{helanderetal2017a}
{\sc \au{Helander, P.}, \au{Parra, F.~I.} \& \au{Newton, S.~L.}}
  \yr{2017{\natexlab{{\em b\/}}}}  \at{Stellarator bootstrap current and plasma
  flow velocity at low collisionality}.  \jt{J. Plasma Phys.}  \bvol{83},
  \pg{905830206}.

\bibitem[Helander \& Sigmar(2002)]{helandersigmar2002}
{\sc \au{Helander, P.} \& \au{Sigmar, D.~J.}} \yr{2002} {\em Collisional
  Transport in Magnetized Plasmas\/}.  \publ{Cambridge University Press}.

\bibitem[Hirsch {\em et~al.\/}(2008)Hirsch, Baldzuhn, Beidler, Brakel, Burhenn,
  Dinklage, Ehmler, Endler, Erckmann, Feng, Geiger, Giannone, Grieger, Grigull,
  Hartfu{\ss}, Hartmann, Jaenicke, K\"{o}nig, Laqua, Maa{\ss}berg, McCormick,
  Sradei, Speth, Stroth, Wagner, Weller, Werner, Wobig \& {S. Zoletnik for the
  W7-AS Team}]{hirschetal2008}
{\sc \au{Hirsch, M.}, \au{Baldzuhn, J.}, \au{Beidler, C.}, \au{Brakel, R.},
  \au{Burhenn, R.}, \au{Dinklage, A.}, \au{Ehmler, H.}, \au{Endler, M.},
  \au{Erckmann, V.}, \au{Feng, Y.}, \au{Geiger, J.}, \au{Giannone, L.},
  \au{Grieger, G.}, \au{Grigull, P.}, \au{Hartfu{\ss}, H.~J.}, \au{Hartmann,
  D.}, \au{Jaenicke, R.}, \au{K\"{o}nig, R.}, \au{Laqua, H.~P.},
  \au{Maa{\ss}berg, H.}, \au{McCormick, K.}, \au{Sradei, F.}, \au{Speth, E.},
  \au{Stroth, U.}, \au{Wagner, F.}, \au{Weller, A.}, \au{Werner, A.},
  \au{Wobig, H.} \& \au{{S. Zoletnik for the W7-AS Team}}} \yr{2008}  \at{Major
  results from the stellarator {Wendelstein 7-AS}}.  \jt{Plasma Phys. Control.
  Fusion}  \bvol{50},  \pg{053001}.

\bibitem[Hirshman(1976)]{hirshman1976}
{\sc \au{Hirshman, S.~P.}} \yr{1976}  \at{Transport properties of a toroidal
  plasma in a mixed collisionality regime}.  \jt{Phys. Fluids}  \bvol{19},
  \pg{155--158}.

\bibitem[Hirshman(1977)]{hirshman1977}
{\sc \au{Hirshman, S.~P.}} \yr{1977}  \at{Transport of a multiple-ion species
  plasma in the {Pfirsch-Schl\"{u}ter} regime}.  \jt{Phys. Fluids}  \bvol{20},
  \pg{589--598}.

\bibitem[Ho \& Kulsrud(1987)]{hokulsrud1987}
{\sc \au{Ho, D.~D.} \& \au{Kulsrud, R.~M.}} \yr{1987}  \at{Neoclassical
  transport in stellarators}.  \jt{Phys. Fluids}  \bvol{30},  \pg{442--461}.

\bibitem[Ida {\em et~al.\/}(2009)Ida, Yoshinuma, Osakabe, Nagaoka, Yokoyama,
  Funaba, Suzuki, Ido, Shimzu, Tamura, Kasahara, Takeiri, Ikeda, Tsumori,
  Kaneko, Morita, Goto, Tanaka, Narihara, Minami, Yamada \& {LHD Experimental
  Group}]{idaetal2009}
{\sc \au{Ida, K.}, \au{Yoshinuma, M.}, \au{Osakabe, M.}, \au{Nagaoka, K.},
  \au{Yokoyama, M.}, \au{Funaba, H.}, \au{Suzuki, C.}, \au{Ido, T.},
  \au{Shimzu, A.}, \au{Tamura, N.}, \au{Kasahara, H.}, \au{Takeiri, Y.},
  \au{Ikeda, K.}, \au{Tsumori, K.}, \au{Kaneko, O.}, \au{Morita, S.}, \au{Goto,
  M.}, \au{Tanaka, K.}, \au{Narihara, K.}, \au{Minami, T.}, \au{Yamada, I.} \&
  \au{{LHD Experimental Group}}} \yr{2009}  \at{Observation of an impurity hole
  in a plasma with an ion internal tranport barrier in the {Large Helical
  Device}}.  \jt{Phys. Plasmas}  \bvol{16},  \pg{056111}.

\bibitem[Igitkhanov {\em et~al.\/}(2006)Igitkhanov, Polunovsky \&
  Beidler]{igitkhanovetal2006}
{\sc \au{Igitkhanov, Y.}, \au{Polunovsky, E.} \& \au{Beidler, C.~D.}} \yr{2006}
   \at{Impurity dynamics in nonaxisymmetric plasmas}.  \jt{Fusion Sci.
  Technol.}  \bvol{50},  \pg{268--275}.

\bibitem[Joffrin {\em et~al.\/}(2014)Joffrin, Baruzzo, Beurskens, Bourdelle,
  Brezinsek, Bucalossi, Buratti, Calabro, Challis, Clever, Coenen, Delabie,
  Dux, Lomas, {de la Luna}, {de Vries}, Flanagan, Frassinetti, Frigione,
  Giroud, Groth, Hawkes, Hobirk, Lehnen, Maddison, Mailloux, Maggi, Matthews,
  Mayoral, Meigs, Neu, Nunes, Puetterich, Rimini, Sertoli, Seiglin, Sips, {van
  Rooij}, Voitsekhovitch \& {JET-EFDA Contributors}]{joffrinetal2014}
{\sc \au{Joffrin, E.}, \au{Baruzzo, M.}, \au{Beurskens, M.}, \au{Bourdelle,
  C.}, \au{Brezinsek, S.}, \au{Bucalossi, J.}, \au{Buratti, P.}, \au{Calabro,
  G.}, \au{Challis, C.~D.}, \au{Clever, M.}, \au{Coenen, J.}, \au{Delabie, E.},
  \au{Dux, R.}, \au{Lomas, P.}, \au{{de la Luna}, E.}, \au{{de Vries}, P.},
  \au{Flanagan, J.}, \au{Frassinetti, L.}, \au{Frigione, D.}, \au{Giroud, C.},
  \au{Groth, M.}, \au{Hawkes, N.}, \au{Hobirk, J.}, \au{Lehnen, M.},
  \au{Maddison, G.}, \au{Mailloux, J.}, \au{Maggi, C.~F.}, \au{Matthews, G.},
  \au{Mayoral, M.}, \au{Meigs, A.}, \au{Neu, R.}, \au{Nunes, I.},
  \au{Puetterich, T.}, \au{Rimini, F.}, \au{Sertoli, M.}, \au{Seiglin, B.},
  \au{Sips, A. C.~C.}, \au{{van Rooij}, G.}, \au{Voitsekhovitch, I.} \&
  \au{{JET-EFDA Contributors}}} \yr{2014}  \at{First scenario development with
  the {JET} new {ITER}-like wall}.  \jt{Nucl. Fusion}  \bvol{54},  \pg{013011}.

\bibitem[Klinger {\em et~al.\/}(2017)Klinger, Alonso, Bozhenkov, Burhenn,
  Dinklage, Fuchert, Geiger, Grulke, Langenberg, Hirsch, Kocsis, Knauer,
  Kr\"{}mer-Flekcen, Laqua, Lazerson, Landreman, Maa{\ss}berg, Marsen, Otte,
  Pablant, Pasch, Rahbarnia, Stange, Szepesi, Thomsen, Traverso, Velasco,
  Wauters, Weir, Windisch \& {The Wendelstein 7-X Team}]{klingeretal2017}
{\sc \au{Klinger, T.}, \au{Alonso, A.}, \au{Bozhenkov, S.}, \au{Burhenn, R.},
  \au{Dinklage, A.}, \au{Fuchert, G.}, \au{Geiger, J.}, \au{Grulke, O.},
  \au{Langenberg, A.}, \au{Hirsch, M.}, \au{Kocsis, G.}, \au{Knauer, J.},
  \au{Kr\"{}mer-Flekcen, A.}, \au{Laqua, H.}, \au{Lazerson, S.}, \au{Landreman,
  M.}, \au{Maa{\ss}berg, H.}, \au{Marsen, S.}, \au{Otte, M.}, \au{Pablant, N.},
  \au{Pasch, E.}, \au{Rahbarnia, K.}, \au{Stange, T.}, \au{Szepesi, T.},
  \au{Thomsen, H.}, \au{Traverso, P.}, \au{Velasco, J.~L.}, \au{Wauters, T.},
  \au{Weir, G.}, \au{Windisch, T.} \& \au{{The Wendelstein 7-X Team}}}
  \yr{2017}  \at{Performance and properties of the first plasmas of wendelstein
  7-x}.  \jt{Plasma Phys. Control. Fusion}  \bvol{59},  \pg{014018}.

\bibitem[Landreman(2017)]{landreman2017}
{\sc \au{Landreman, M.}} \yr{2017}  \at{An improved current potential method
  for fast computation of stellarator coil shapes}.  \jt{Nucl. Fusion}
  \bvol{57},  \pg{046003}.

\bibitem[Landreman {\em et~al.\/}(2014)Landreman, Smith, Moll\'{e}n \&
  Helander]{landremanetal2014}
{\sc \au{Landreman, M.}, \au{Smith, H.~M.}, \au{Moll\'{e}n, A.} \&
  \au{Helander, P.}} \yr{2014}  \at{Comparison of particle trajectories and
  collision operators for collisional transport in nonaxisymmetric plasmas}.
  \jt{Phys. Plasmas}  \bvol{21},  \pg{042503}.

\bibitem[Moll\'{e}n {\em et~al.\/}(2015)Moll\'{e}n, Landreman, Smith, Braun \&
  Helander]{mollenetal2015}
{\sc \au{Moll\'{e}n, A.}, \au{Landreman, M.}, \au{Smith, H.~M.}, \au{Braun, S.}
  \& \au{Helander, P.}} \yr{2015}  \at{Impurities in a non-axisymmetric plasma:
  tranport and effect on bootstrap current}.  \jt{Phys. Plasmas}  \bvol{22},
  \pg{112508}.

\bibitem[Nakajima {\em et~al.\/}(1989)Nakajima, Okamoto, Todoroki, Nakamura \&
  Wakatani]{nakajimaetal1989}
{\sc \au{Nakajima, N.}, \au{Okamoto, M.}, \au{Todoroki, J.}, \au{Nakamura, Y.}
  \& \au{Wakatani, M.}} \yr{1989}  \at{Optimization of the bootstrap current in
  a large helical system with {L=2}}.  \jt{Nucl. Fusion}  \bvol{29},
  \pg{605--616}.

\bibitem[Newton \& Helander(2006)]{newtonhelander2006}
{\sc \au{Newton, S.} \& \au{Helander, P.}} \yr{2006}  \at{Neoclassical momentum
  transport in an impure rotating tokamak plasma}.  \jt{Phys. Plasmas}
  \bvol{13},  \pg{012505}.

\bibitem[Rosenbluth {\em et~al.\/}(1972)Rosenbluth, Hazeltine \&
  Hinton]{rosenbluthetal1972}
{\sc \au{Rosenbluth, M.~N.}, \au{Hazeltine, R.~D.} \& \au{Hinton, F.~L.}}
  \yr{1972}  \at{Plasma transport in toroidal confinement systems}.  \jt{Phys.
  Fluids}  \bvol{15},  \pg{116--140}.

\bibitem[Rutherford(1974)]{rutherford1974}
{\sc \au{Rutherford, P.~H.}} \yr{1974}  \at{Impurity transport in the
  {Pfirsch-Schl\"{u}ter} regime}.  \jt{Phys. Fluids}  \bvol{17},  \pg{1782}.

\bibitem[Samain \& Werkoff(1977)]{samainwerkoff1977}
{\sc \au{Samain, A.} \& \au{Werkoff, F.}} \yr{1977}  \at{Diffusion in tokamaks
  with impurities in the {Pfirsch-Schl\"{u}ter} regime}.  \jt{Nucl. Fusion}
  \bvol{17},  \pg{53--64}.

\bibitem[Sugama \& Nishimura(2002)]{sugamanishimura2002}
{\sc \au{Sugama, H.} \& \au{Nishimura, S.}} \yr{2002}  \at{How to calculate the
  neoclassical viscosity, diffusion, and current coefficients in general
  toroidal plasmas}.  \jt{Phys. Plasmas}  \bvol{9},  \pg{4637--4653}.

\bibitem[Velasco {\em et~al.\/}(2017)Velasco, Calvo, Satake, Alonso, Nunami,
  Yokoyama, Sato, Estrada, Fontdecaba, Liniers, McCarthy, Medina, {Ph Van
  Milligen}, Ochando, Parra, Sugama, Zhezhera, {The LHD Experimental Team} \&
  {The TJ-II Team}]{velascoetal2017}
{\sc \au{Velasco, J.~L.}, \au{Calvo, I.}, \au{Satake, S.}, \au{Alonso, A.},
  \au{Nunami, M.}, \au{Yokoyama, M.}, \au{Sato, M.}, \au{Estrada, T.},
  \au{Fontdecaba, J.~M.}, \au{Liniers, M.}, \au{McCarthy, K.~J.}, \au{Medina,
  F.}, \au{{Ph Van Milligen}, B.}, \au{Ochando, M.}, \au{Parra, F.},
  \au{Sugama, H.}, \au{Zhezhera, A.}, \au{{The LHD Experimental Team}} \&
  \au{{The TJ-II Team}}} \yr{2017}  \at{Moderation of neoclassical impurity
  accumulation in high temperature plasmas of helical devices}.  \jt{Nucl.
  Fusion}  \bvol{57},  \pg{016016}.

\bibitem[{W VII-A Team} \& {NI Group}(1985)]{w7a1985}
{\sc \au{{W VII-A Team}} \& \au{{NI Group}}} \yr{1985}  \at{Impurity transport
  in the {Wendelstein VII-A} stellarator}.  \jt{Nucl. Fusion}  \bvol{25},
  \pg{1593--1609}.

\bibitem[Wade {\em et~al.\/}(2000)Wade, Houlberg \& Baylor]{wadeetal2000}
{\sc \au{Wade, M.~R.}, \au{Houlberg, W.~A.} \& \au{Baylor, L.~R.}} \yr{2000}
  \at{Experimental confirmation of impurity convection driven by the
  ion-temperature gradient in toroidal plasmas}.  \jt{Phys. Rev. Lett.}
  \bvol{84},  \pg{282--285}.

\end{thebibliography}

\end{document}